\documentclass[pdflatex,sn-mathphys-num]{sn-jnl}

\usepackage{graphicx}%
\usepackage{multirow}%
\usepackage{amsmath,amssymb,amsfonts}%
\usepackage{amsthm}%
\usepackage{mathrsfs}%
\usepackage[title]{appendix}%
\usepackage{xcolor}%
\usepackage{textcomp}%
\usepackage{manyfoot}%
\usepackage{booktabs}%
\usepackage{algorithm}%
\usepackage{algorithmicx}%
\usepackage{algpseudocode}%
\usepackage{listings}%
\usepackage{url}%

\theoremstyle{thmstyleone}%
\theoremstyle{thmstyletwo}%

\theoremstyle{thmstylethree}%

\begin{document}

\title[Technological Fitness and Regional Growth in Japan]{Technological Fitness and Regional Economic Growth in Japanese Prefectures}

\author[1]{\fnm{Rintaro} \sur{Karashima}}
\equalcont{These authors contributed equally to this work.}

\author*[1,2]{\fnm{Hiroyasu} \sur{Inoue}}
\email{inoue@gsis.u-hyogo.ac.jp}
\equalcont{These authors contributed equally to this work.}

\affil*[1]{\orgdiv{Graduate School of Information Science}, \orgname{University of Hyogo}, \orgaddress{\street{7-1-28 Minatojima-minamimachi, Chuo-ku}, \city{Kobe}, \postcode{650-0047}, \state{Hyogo}, \country{Japan}}}

\affil[2]{\orgdiv{Center for Computational Science}, \orgname{RIKEN}, \orgaddress{\street{7-1-26 Minatojima-minamimachi, Chuo-ku}, \city{Kobe}, \postcode{650-0047}, \state{Hyogo}, \country{Japan}}}

\abstract{%
Technological knowledge plays an important role in shaping regional economic performance.
This study examines the relationship between the sophistication of regional technological capabilities and economic growth across Japanese prefectures.
Using approximately 3.9 million corporate patent records filed from fiscal years 1981 to 2015, we construct bipartite networks linking 47 prefectures to 35 technology classes and apply the Fitness-Complexity algorithm to derive regional Fitness scores for seven five-year periods.
We estimate fixed-effects panel models with Driscoll-Kraay standard errors, using the annual average growth rate of real gross regional product per capita over the subsequent five years as the dependent variable.
Prefectural Fitness is positively associated with subsequent growth ($\hat{\beta} = 0.0029$, $p = 0.007$) after controlling for initial income, population density, and patenting activity, but this relationship is detectable only when both entity and time fixed effects are included.
Cross-sectional correlations between Fitness and subsequent growth change sign across periods, underscoring the importance of the panel approach.
The growth effect of Fitness is stronger in prefectures with lower initial income, suggesting that technological sophistication contributes more to growth where there is greater scope for economic expansion.
Lag and lead analyses indicate that the relationship runs from Fitness to subsequent growth rather than the reverse.%
}


\maketitle

\section{Introduction}\label{section:Introduction}

Since the late twentieth century, the production of technological knowledge has been recognized as one of the key factors shaping the economic performance of nations and regions \cite{hidalgo2009eci,Hidalgo2021,Balland2022}.
Understanding how the structure of technological capabilities relates to growth therefore matters for regional innovation policy and for the design of evidence-based strategies \cite{Hidalgo2023,Bahrami2023}.

One influential approach to measuring these capabilities is the method of reflections proposed by Hidalgo and Hausmann \cite{hidalgo2009eci}.
This method represents the relationship between economic actors and the activities they perform as a bipartite network, and derives complexity indicators through iterative computation on this network.
The resulting index assigns higher values to actors that are connected to rare and sophisticated activities, which in turn are those practiced by few but diversified actors.
Applications of this framework have shown that countries with higher complexity tend to achieve higher income growth over subsequent years \cite{Hidalgo2021,Hidalgo2023}.

The Fitness-Complexity (FC) algorithm \cite{Tacchella2012} is a nonlinear refinement of this approach that addresses convergence issues in the original linear method \cite{Albeaik2017}.
At each iteration, Fitness is updated as the sum of the complexities of connected activities, while complexity is updated as the inverse of the mean inverse-Fitness of connected actors.
This nonlinear structure penalizes activities practiced by low-Fitness actors more strongly than linear averaging, which makes the FC algorithm more robust when the bipartite network is sparse.
Empirical studies have confirmed that the resulting Fitness index is a useful predictor of growth at the national level \cite{morrison2017economic} and for cities using patent data \cite{straccamore2023urban,straccamore2025comparative}.

Applying this framework to patent data at the regional level has attracted increasing attention.
Balland and Rigby \cite{balland2017tci} introduced the Technological Complexity Index using US metropolitan-area patents, and subsequent work has extended regional patent-based complexity analysis to European regions \cite{PintarScherngell2022,Pinheiro2022,ballandandboschma2021}, Ireland \cite{Whittle2019}, Korea \cite{Jun2023}, and Turkey \cite{Abay2024}.
These studies generally find that regions with more sophisticated technological portfolios tend to grow faster and are better positioned to diversify into new technological domains.
However, published applications to Japanese prefectures remain limited.
Chakraborty et al.\ \cite{Chakraborty2020} applied the Hidalgo-Hausmann algorithm to prefecture-level industrial activity data but focused on trade-based economic complexity rather than patent-based technological sophistication.

This study fills this gap by applying the FC algorithm to Japanese prefectural patent data and examining whether the resulting Fitness index predicts subsequent regional growth.
Using approximately 3.9 million corporate patents filed from fiscal years 1981 to 2015, we construct bipartite networks for 47 prefectures over seven five-year periods, derive Fitness scores, and estimate fixed-effects panel models with Driscoll-Kraay standard errors.
The analysis yields three findings.
First, prefectural Fitness is a significant positive predictor of subsequent growth, but only when both entity and time fixed effects are controlled for.
Cross-sectional correlations between Fitness and growth change sign across periods, which demonstrates that naive regressions without adequate controls can be misleading.
Second, the growth effect of Fitness is stronger in lower-income prefectures, consistent with the view that regions with greater scope for catch-up benefit more from accumulating sophisticated technological capabilities.
Third, causality checks using lagged and led Fitness values indicate that the relationship runs from Fitness to growth rather than the reverse.

The rest of this paper is organized as follows.
The Results section presents the regional Fitness distribution, panel estimation results, heterogeneous effects by income level, and causality checks.
The Discussion section interprets the findings and discusses limitations.
The Methods section describes the data, network construction, Fitness estimation, and panel model specification.

\section{Results}\label{section:Results}

\subsection{Regional Technological Characteristics}\label{subsection:characteristics}

\begin{figure}[ht]
  \includegraphics[scale=0.6]{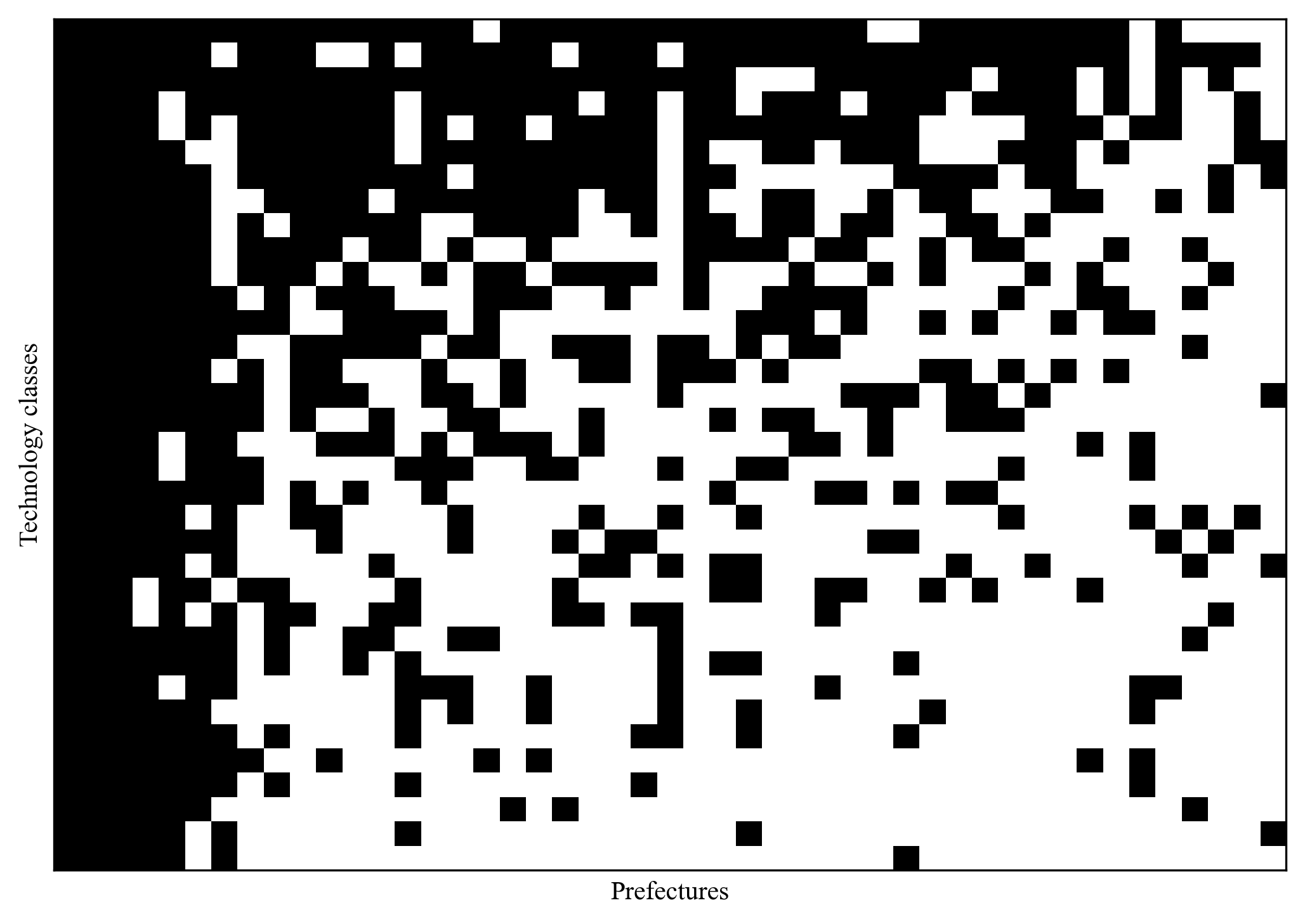}
  \caption{%
    Adjacency matrix of the bipartite network connecting prefectures and technology classes.
    Rows represent technology classes sorted from top to bottom by increasing rarity, and columns represent prefectures sorted from left to right by decreasing diversity.
    An entry of one indicates that the prefecture holds a technological advantage in that class.
    The approximately triangular concentration of ones in the upper-left region reflects the nested structure of the network.%
  }
  \label{fig:bipartite}
\end{figure}

The bipartite adjacency matrix constructed from the patent data exhibits a nested structure, in which high-diversity prefectures hold links to both common and rare technological classes, while low-diversity prefectures concentrate on more common classes (Fig.~\ref{fig:bipartite}).
This pattern is consistent with findings from regional patent-based complexity studies in other countries \cite{balland2017tci,PintarScherngell2022} and reflects the idea that the production of rare technologies requires a broad set of capabilities that is also sufficient for producing common ones \cite{hidalgo2009eci}.

\begin{figure}[ht]
  \includegraphics[scale=0.6]{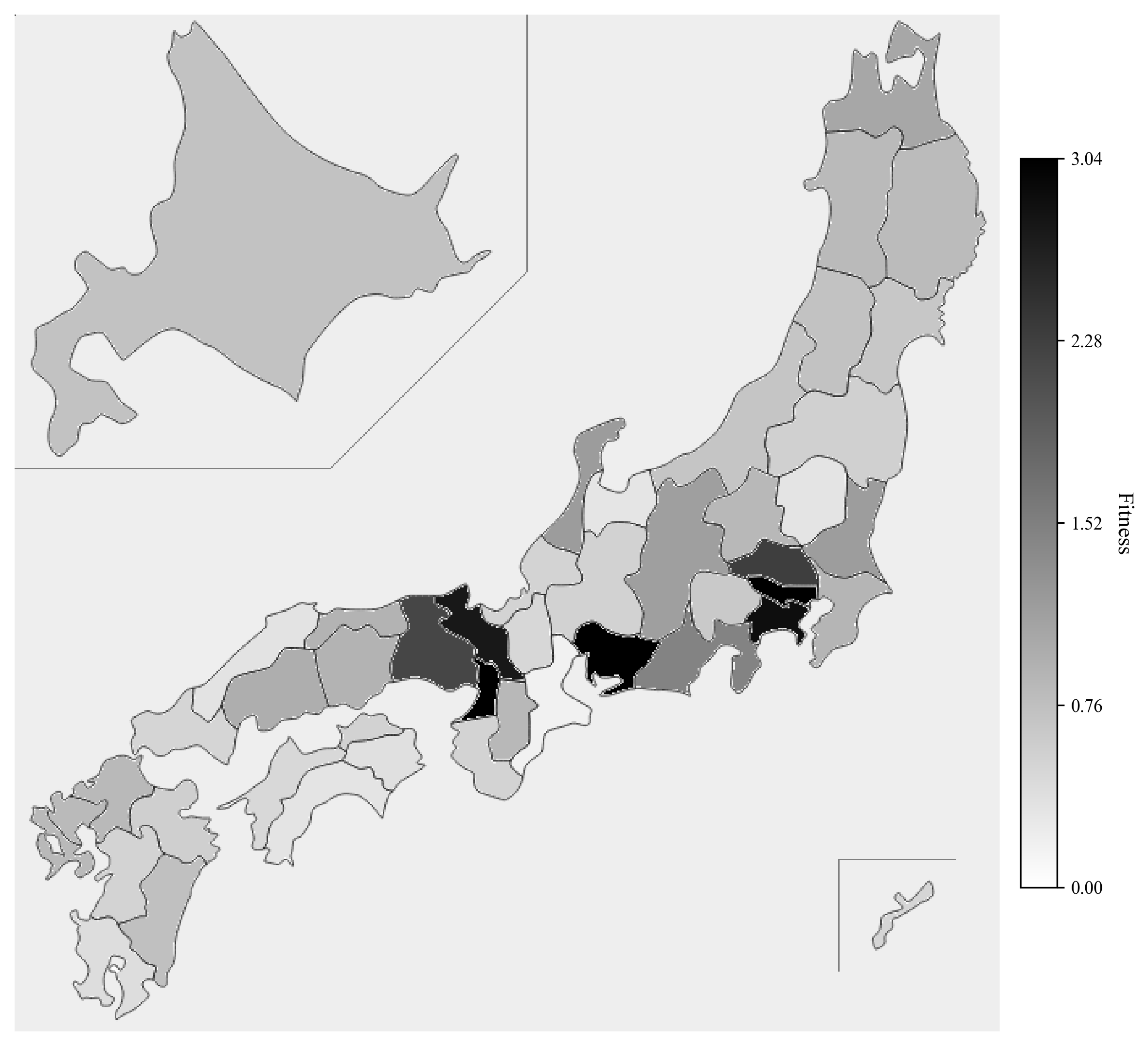}
  \caption{%
    Prefectural Fitness values for the 2011--2015 period displayed as a choropleth map of Japan.
    Darker shading indicates higher Fitness.
    Prefectures in the three major industrial corridors, Kanto, Tokai, and Kinki, show the highest values.%
  }
  \label{fig:fitnessmap}
\end{figure}

Focusing on diversity, prefectures belonging to Japan's three major metropolitan areas and their surrounding regions show the broadest technological portfolios.
Prefectures in the Keihanshin area, which includes Osaka and Kyoto, and in the Keihin area, which includes Tokyo and Kanagawa, hold links to both common classes such as Food Chemistry and Biotechnology and to rare classes such as Computer Technology and Digital Communication.
These patterns translate into high Fitness values, as illustrated in the prefectural map (Fig.~\ref{fig:fitnessmap}).

Table~\ref{tab:fitnessranking} reports the Fitness values and rankings for all 47 prefectures in the 2011--2015 period.
Osaka, Aichi, and Tokyo share the top rank with a Fitness of 3.04, followed by Kanagawa at 2.87 and Kyoto at 2.73.
Saitama and Hyogo rank sixth and seventh with Fitness values of 2.29 and 2.20, respectively.
These results confirm that high Fitness is concentrated in the major industrial corridors of central Japan.
At the same time, several prefectures outside the major metropolitan areas, including Ishikawa and Ibaraki, achieve relatively high Fitness despite narrower technological portfolios.
In these cases, links to rare fields such as Telecommunications and Control appear to support high Fitness even without broad diversity, which is consistent with the structure of the FC algorithm, where rare-technology connections contribute substantially to the Fitness score.

\begin{table}[ht]
\caption{Fitness rankings for all 47 prefectures in the 2011--2015 period}\label{tab:fitnessranking}
\small
\begin{tabular}{rlc @{\hspace{1.5em}} rlc @{\hspace{1.5em}} rlc}
\toprule
Rank & Prefecture & Fitness & Rank & Prefecture & Fitness & Rank & Prefecture & Fitness \\
\midrule
 1 & Osaka     & 3.04 & 17 & Nara      & 0.85 & 33 & Kagawa    & 0.54 \\
 1 & Tokyo     & 3.04 & 18 & Gunma     & 0.85 & 34 & Wakayama  & 0.53 \\
 1 & Aichi     & 3.04 & 19 & Nagasaki  & 0.84 & 35 & Fukui     & 0.52 \\
 4 & Kanagawa  & 2.87 & 20 & Fukuoka   & 0.83 & 36 & Kumamoto  & 0.51 \\
 5 & Kyoto     & 2.73 & 21 & Saga      & 0.82 & 37 & Okinawa   & 0.51 \\
 6 & Saitama   & 2.29 & 22 & Akita     & 0.82 & 38 & Yamaguchi & 0.50 \\
 7 & Hyogo     & 2.20 & 23 & Iwate     & 0.82 & 39 & Shiga     & 0.48 \\
 8 & Shizuoka  & 1.47 & 24 & Miyazaki  & 0.75 & 40 & Ehime     & 0.47 \\
 9 & Ishikawa  & 1.18 & 25 & Hokkaido  & 0.73 & 41 & Kagoshima & 0.40 \\
10 & Ibaraki   & 1.17 & 26 & Yamagata  & 0.73 & 42 & Tokushima & 0.38 \\
11 & Nagano    & 1.15 & 27 & Niigata   & 0.70 & 43 & Shimane   & 0.34 \\
12 & Aomori    & 1.05 & 28 & Miyagi    & 0.67 & 44 & Tochigi   & 0.33 \\
13 & Hiroshima & 0.97 & 29 & Yamanashi & 0.65 & 45 & Kochi     & 0.32 \\
14 & Okayama   & 0.91 & 30 & Oita      & 0.57 & 46 & Toyama    & 0.32 \\
15 & Tottori   & 0.90 & 31 & Fukushima & 0.57 & 47 & Mie       & 0.18 \\
16 & Chiba     & 0.89 & 32 & Gifu      & 0.55 &    &           &      \\
\bottomrule
\end{tabular}
\end{table}

\subsection{Fitness and Regional Economic Growth}\label{subsection:growth}

Before estimating the panel model, we verify that multicollinearity among the explanatory variables is not a concern by computing variance inflation factors (VIF) after applying the within-group transformation.
All VIF values fall below 1.15, well below the conventional threshold of 10, confirming that multicollinearity does not affect the estimates (see Appendix~\ref{secA1}).

To motivate the panel approach, we first examine the cross-sectional relationship between Fitness and subsequent five-year growth for each period separately.
Table~\ref{tab:crosssectional} reports Pearson correlation coefficients and their significance levels.
The sign of the correlation changes across periods, from positive and insignificant in 1986 ($r = 0.188$, $p = 0.205$) to negative and significant in 1991 ($r = -0.390$, $p = 0.007$), then to positive and significant in 2001 ($r = 0.442$, $p = 0.002$), and back to negative and significant in 2016 ($r = -0.383$, $p = 0.008$).
This instability indicates that the simple cross-sectional relationship is heavily confounded by time-varying shocks and time-invariant regional characteristics, and that a panel model with fixed effects is necessary to identify the relationship of interest.

\begin{table}[ht]
\caption{Cross-sectional Pearson correlations between Fitness and the subsequent five-year growth rate of real GRP per capita}\label{tab:crosssectional}
\begin{tabular}{cccc}
\toprule
Base year ($\tau$) & $n$ & Correlation ($r$) & $p$-value \\
\midrule
1986 & 47 & $+0.188$ & 0.205  \\
1991 & 47 & $-0.390$ & 0.007** \\
1996 & 47 & $-0.228$ & 0.123  \\
2001 & 47 & $+0.442$ & 0.002** \\
2006 & 47 & $-0.248$ & 0.093  \\
2011 & 47 & $-0.281$ & 0.055  \\
2016 & 47 & $-0.383$ & 0.008** \\
\bottomrule
\multicolumn{4}{l}{$^{**}p < 0.01$.}\\
\end{tabular}
\end{table}

We estimate four panel specifications that vary the combination of fixed effects.
Model 1 includes no fixed effects, Model 2 adds entity fixed effects only, Model 3 adds time fixed effects only, and Model 4 includes both entity and time fixed effects.
In all models, the dependent variable is the annual average log growth rate of real GRP per capita over the subsequent five years, and the controls are the log of initial GRP per capita, log of population density, and log of patent count.
Standard errors follow Driscoll-Kraay \cite{DriscollandKraay1998} with a bandwidth of three to account for serial and cross-sectional dependence across the seven periods.

Table~\ref{tab:models} reports the Fitness coefficient across the four specifications.
The coefficient is small and statistically indistinguishable from zero in Models 1 through 3.
In Model 4, which includes both sets of fixed effects, the coefficient is positive and significant ($\hat{\beta} = 0.0029$, $t = 2.72$, $p = 0.007$, within $R^2 = 0.694$).
This pattern mirrors the instability seen in the cross-sectional correlations and indicates that both entity and time fixed effects are needed to identify the growth effect of Fitness.

\begin{table}[ht]
\caption{Fitness coefficient estimates under alternative fixed-effects specifications}\label{tab:models}
\begin{tabular}{lcccc}
\toprule
 & Model 1 & Model 2 & Model 3 & Model 4 \\
 & (No FE) & (Entity FE) & (Time FE) & (Two-way FE) \\
\midrule
Fitness    & 0.001   & 0.002  & $-0.000$ & $0.003^{**}$ \\
           & $(0.43)$ & $(0.90)$ & $(-0.12)$ & $(2.72)$ \\
\midrule
Within $R^2$ &       &        &          & 0.694 \\
$N$          & 329   & 329    & 329      & 329 \\
Controls     & Yes   & Yes    & Yes      & Yes \\
Entity FE    & No    & Yes    & No       & Yes \\
Time FE      & No    & No     & Yes      & Yes \\
\bottomrule
\multicolumn{5}{l}{$t$-statistics in parentheses. $^{**}p < 0.01$.}\\
\multicolumn{5}{l}{Controls include $\ln(\text{GRP pc})$, $\ln(\text{capita density})$, and $\ln(\text{patent count})$.}\\
\multicolumn{5}{l}{Driscoll-Kraay standard errors with bandwidth = 3.}\\
\end{tabular}
\end{table}

The full coefficient estimates for Model 4 are reported in Table~\ref{tab:fullmodel} in the Supplementary Information.
The log of initial GRP per capita carries a significant negative coefficient ($\hat{\beta} = -0.113$, $p < 0.001$), consistent with a within-prefecture convergence effect in which periods of higher initial income are followed by slower growth.
Log population density enters with a negative coefficient ($\hat{\beta} = -0.059$, $p = 0.001$), which within the two-way fixed-effects framework captures the association between density increases above a prefecture's own average and subsequent growth.
Log patent count has a positive coefficient ($\hat{\beta} = 0.003$, $p = 0.002$), indicating that periods with higher patenting activity predict somewhat higher growth.

\subsection{Heterogeneous Effects by Income Level}\label{subsection:heterogeneous}

To examine whether the growth effect of Fitness varies with initial income, we augment the baseline specification with an interaction term between Fitness and the log of initial GRP per capita.
The interaction coefficient is negative and significant ($\hat{\beta}_{\text{int}} = -0.0114$, $t = -3.66$, $p < 0.001$), while the main Fitness coefficient is positive and significant ($\hat{\beta}_{\text{Fitness}} = 0.174$, $t = 3.72$, $p < 0.001$).
These estimates imply that the marginal growth effect of Fitness is $0.174 - 0.0114 \times \ln(\text{GRP pc})$, which is larger for prefectures with lower initial income.

This result is consistent with the view that the returns to technological sophistication are higher in less-developed regions where there is greater potential for catch-up growth.
Prefectures already near the technological frontier may benefit less at the margin from further improvements in Fitness, while prefectures further from the frontier can leverage accumulated technological capabilities to open new production possibilities that were previously inaccessible.

\subsection{Causality Checks}\label{subsection:causality}

To assess whether the observed relationship reflects the forward-looking predictive power of Fitness rather than reverse causality, we perform two sets of checks.
First, we replace contemporaneous Fitness with the one-period lagged value, corresponding to Fitness computed for the period five years earlier.
Second, we replace it with future values, using Fitness one and two periods ahead.

The lagged Fitness specification yields a positive and significant coefficient ($\hat{\beta}_{\text{lag1}} = 0.0039$, $t = 6.71$, $p < 0.001$), larger in magnitude than the contemporaneous estimate.
This finding indicates that technological sophistication accumulated in prior periods contributes to subsequent growth, and that the predictive content of Fitness is not limited to contemporaneous observations.

The lead specifications produce coefficients that are not statistically significant ($\hat{\beta}_{\text{lead1}} = 0.0018$, $t = 1.23$, $p = 0.220$; $\hat{\beta}_{\text{lead2}} = -0.0065$, $t = -1.86$, $p = 0.064$).
The absence of significant lead effects indicates that current growth rates do not predict future Fitness, which argues against the interpretation that higher growth mechanically produces higher future technological sophistication.
Taken together, the lag and lead results support a causal channel running from Fitness to subsequent growth rather than the reverse.

\section{Discussion}\label{section:Discussion}

This study provides evidence that prefectural Fitness is positively associated with subsequent economic growth across Japanese prefectures, with a coefficient of 0.0029 per unit of Fitness under the two-way fixed-effects specification.
This finding adds to the growing evidence that patent-based complexity measures carry information about future regional economic performance \cite{balland2017tci,mewes2022technological,LiRigby2023}.

A key observation is that the significant positive association between Fitness and growth is detectable only when both entity and time fixed effects are controlled for.
Without both sets of fixed effects, the estimated coefficient is small and indistinguishable from zero.
The cross-sectional correlations change sign across periods, and this instability confirms that the simple cross-sectional relationship is heavily confounded.
These results highlight the importance of panel methods with appropriate fixed effects when estimating the relationship between complexity indicators and regional growth, a point that has received increasing attention in the complexity literature \cite{mewes2022technological,Pinheiro2022}.

The heterogeneity finding, that the growth effect of Fitness is stronger in lower-income prefectures, is consistent with a conditional convergence framework in which technological sophistication facilitates catch-up growth but generates smaller returns once a region is already near the frontier \cite{Hidalgo2021}.
This pattern has analogues in the growth literature where complementarities between capabilities and the distance from the frontier generate heterogeneous returns to capability accumulation.
In the present context, prefectures such as Osaka, Aichi, and Tokyo are already at the top of the Fitness distribution, so further improvements in Fitness contribute less to growth at the margin.
For prefectures further down the distribution, moving toward a more sophisticated technological portfolio appears to support more substantial subsequent growth.

The causality checks provide additional support for the interpretation that Fitness is an input to growth rather than an output of it.
The lagged specification yields a larger and more precisely estimated coefficient than the contemporaneous one, consistent with the idea that technological sophistication takes time to translate into economic outcomes.
The insignificant lead coefficients indicate that growth does not precede future Fitness, reducing concern that the estimated relationship reflects reverse causality.

Several limitations should be noted.
First, the analysis uses a single algorithm with a specific link rule and classification system.
Robustness to alternative complexity measures, such as the Economic Complexity Index or citation-weighted patent variants, remains to be investigated \cite{mewes2022technological,PintarEssletzbichler2022}.
Second, the identification strategy relies on two-way fixed effects, which removes unobserved time-invariant regional characteristics and common time trends but does not address time-varying confounders.
Variables such as regional research and development investment, human capital, and industrial structure changes may simultaneously influence Fitness and growth.
Third, the five-year aggregation windows smooth out short-run fluctuations but may also average over economically meaningful within-period dynamics.
Fourth, the analysis is limited to Japan, and it is not clear whether the observed relationships hold in other institutional settings where the patent system, industrial structure, or regional governance differs \cite{straccamore2025comparative}.

Future work can extend the analysis to alternative complexity measures and incorporate citation-based patent quality indicators for robustness \cite{PintarScherngell2022}.
Adding controls for human capital and research and development investment could sharpen identification.
Comparative analyses across countries with rich regional patent data would test whether the role of Fitness in regional growth is specific to Japan or more general \cite{straccamore2025comparative}.

\section{Methods}\label{section:Methods}

\subsection{Data}\label{subsection:Data}

The patent data used in this study come from the Japan Patent Office.
The dataset covers approximately 3.9 million registered patents filed by corporations from fiscal years 1981 to 2015.
Each patent record contains information on the applicant's address, which is used to assign the patent to a prefecture, and the International Patent Classification (IPC) codes.
The primary IPC code of each patent is mapped to one of the 35 technology classes defined by Schmoch \cite{Schmoch2008}.
The Schmoch classification reorganizes IPC codes by considering the homogeneity of technologies and the balance across class sizes, and it has been widely applied in regional patent-based complexity analyses \cite{PintarScherngell2022,Balland2018,Whittle2019}.
One advantage of using this classification for long-run analyses is that it accounts for historical changes in patent categories and enables consistent comparisons across decades.

A single patent may involve multiple applicants located in different prefectures or may be associated with multiple primary IPC codes in the case of complex technologies or joint applications.
Direct counting in these cases would overestimate the patent activity of individual prefectures and technology classes.
Following the approach validated by Pintar et al.\ \cite{PintarScherngell2022}, we fractionally allocate each patent by dividing its contribution equally among the linked prefectures and technology classes.
For example, a patent involving two applicants in different prefectures and two primary IPC codes contributes 0.25 to each prefecture-technology pair.

The economic outcome variable is the real Gross Regional Product (GRP) per capita for each prefecture.
Real GRP measures the total value added produced within a prefecture, and the per-capita measure is obtained by dividing real GRP by the resident population.
GRP and population data for fiscal years 1981 to 2020 were obtained from the Statistics Bureau of Japan via the e-Stat portal (\url{https://www.e-stat.go.jp}).

\subsection{Bipartite Network Construction}\label{subsection:bipartite}

To determine whether a prefecture holds a technological advantage in a given class, we compute the Revealed Technological Advantage (RTA) index \cite{Soete1987}, which adapts the revealed comparative advantage concept of Balassa \cite{Balassa1965} to patent data.
Let $X_{pc}$ denote the fractional patent count of prefecture $p$ in technology class $c$.
The RTA index is defined as

\begin{equation}\label{eq:rta}
\mathrm{RTA}_{pc} = \frac{X_{pc} / \sum_{c} X_{pc}}{\sum_{p} X_{pc} / \sum_{p}\sum_{c} X_{pc}}.
\end{equation}

The standard approach assigns a link when $\mathrm{RTA}_{pc} \geq 1$.
However, this threshold tends to underrepresent large metropolitan prefectures whose share of any individual technology class appears small relative to their overall size \cite{FritzManduca2021,PintarScherngell2022}.
To address this, we follow Fritz and Manduca \cite{FritzManduca2021} and Pintar et al.\ \cite{PintarScherngell2022} and apply a composite condition: a link is assigned when either $\mathrm{RTA}_{pc} \geq 1$ or $X_{pc} \geq a_c$, where $a_c$ is a technology-class-specific threshold based on the distribution of patent counts in class $c$.
The binary adjacency matrix $M_{pc}$ is therefore

\begin{equation}\label{eq:adjacency}
M_{pc} =
\begin{cases}
1 & \text{if } \mathrm{RTA}_{pc} \geq 1 \text{ or } X_{pc} \geq a_c, \\
0 & \text{otherwise.}
\end{cases}
\end{equation}

The degree centrality $\sum_{c} M_{pc}$ of prefecture $p$ in the bipartite network represents its diversity, and the inverse degree centrality $1/\sum_{p} M_{pc}$ of technology class $c$ represents its rarity.

\subsection{Fitness-Complexity Algorithm}\label{subsection:fitness}

The FC algorithm \cite{Tacchella2012} computes the sophistication of each prefecture and the complexity of each technology class through nonlinear iterative updates.
Let $F_{p}^{(N)}$ denote the Fitness of prefecture $p$ and $Q_{c}^{(N)}$ the complexity of technology class $c$ at iteration $N$.
Starting from $F_{p}^{(0)} = Q_{c}^{(0)} = 1$, the updates are

\begin{equation}\label{eq:fitness}
\widetilde{F}_{p}^{(N)} = \sum_{c} M_{pc}\, Q_{c}^{(N-1)}, \qquad
F_{p}^{(N)} = \frac{\widetilde{F}_{p}^{(N)}}{\langle \widetilde{F}_{p}^{(N)} \rangle},
\end{equation}

\begin{equation}\label{eq:complexity}
\widetilde{Q}_{c}^{(N)} = \left[ \sum_{p} M_{pc} \left( F_{p}^{(N-1)} \right)^{-1} \right]^{-1}, \qquad
Q_{c}^{(N)} = \frac{\widetilde{Q}_{c}^{(N)}}{\langle \widetilde{Q}_{c}^{(N)} \rangle},
\end{equation}

where $\langle \cdot \rangle$ denotes the mean over the indicated nodes.
Fitness at each step is proportional to the average complexity of the technologies a prefecture practices, while complexity at each step is inversely related to the average Fitness of the prefectures that practice it.
This nonlinear structure penalizes activities practiced by low-Fitness actors more strongly than linear averaging, making the FC algorithm more robust in sparse networks compared with the method of reflections \cite{Tacchella2012,Albeaik2017}.

We assess convergence using the Minimum Crossing Iteration criterion \cite{Pugliese2016}, which terminates when the number of iterations until any rank order change exceeds $10^{5}$.
The Fitness index is computed separately for each of the seven five-year periods: 1981--1985, 1986--1990, 1991--1995, 1996--2000, 2001--2005, 2006--2010, and 2011--2015.
Using five-year periods rather than individual years reduces fluctuations in estimated Fitness values that arise when network density varies strongly across single years \cite{straccamore2023urban}.

\subsection{Panel Estimation}\label{subsection:estimation}

The panel model relates the annual average growth rate of real GRP per capita over the subsequent five years to current Fitness and a set of controls.
The dependent variable for period $t$ is

\begin{equation}\label{eq:growth}
g_{p,t} = \frac{1}{5} \left[ \ln \mathrm{GRPpc}_{p,t+5} - \ln \mathrm{GRPpc}_{p,t} \right],
\end{equation}

which measures the average annual log growth rate over the five years following the Fitness measurement period.
The baseline specification is

\begin{equation}\label{eq:panel}
g_{p,t} = \alpha + \beta_{1} \ln \mathrm{GRPpc}_{p,t} + \beta_{2} \mathrm{Fitness}_{p,t}
         + \beta_{3} \ln \mathrm{density}_{p,t} + \beta_{4} \ln \mathrm{Patent}_{p,t}
         + \mu_{p} + \lambda_{t} + \varepsilon_{p,t},
\end{equation}

where $\ln \mathrm{GRPpc}_{p,t}$ is the log of real GRP per capita at the beginning of period $t$, $\mathrm{Fitness}_{p,t}$ is the Fitness score for period $t$, $\ln \mathrm{density}_{p,t}$ is the log of population density, $\ln \mathrm{Patent}_{p,t}$ is the log of total patent count during period $t$, $\mu_{p}$ is a prefecture fixed effect, $\lambda_{t}$ is a period fixed effect, and $\varepsilon_{p,t}$ is the error term.
The panel covers $N = 47 \times 7 = 329$ observations.

Prefecture fixed effects absorb time-invariant regional characteristics such as geographic location and industrial history.
Period fixed effects absorb macroeconomic shocks common to all prefectures, such as national policy changes and cyclical fluctuations.
Without both sets of fixed effects, the estimated coefficient on Fitness may reflect these confounders rather than the direct relationship between technological sophistication and growth.

The error term $\varepsilon_{p,t}$ may be serially correlated across periods within a prefecture and spatially correlated across prefectures within a period.
Ordinary least squares standard errors would be downward biased in this case even if the coefficient estimates are consistent.
We therefore use Driscoll-Kraay standard errors \cite{DriscollandKraay1998}, which are asymptotically valid in the presence of both time-series dependence and cross-sectional dependence.
Given the short panel dimension of seven periods, we set the bandwidth to three to ensure stable estimation.

To examine heterogeneous effects by income level, we extend the baseline model with an interaction term between Fitness and $\ln \mathrm{GRPpc}_{p,t}$.
To investigate causality, we replace contemporaneous Fitness with the lagged value from the previous period or with lead values from one or two periods ahead.
The lead specification tests whether current growth predicts future Fitness, which would indicate reverse causality.

\backmatter

\bmhead{Supplementary information}

Two supplementary figures and one supplementary table are provided in Appendix~\ref{secA1}.
Fig.~S1 shows variance inflation factors for all regressors after the within-group transformation.
Fig.~S2 shows scatter plots of Fitness against subsequent five-year growth for each period.
Table~S1 reports the full coefficient estimates for the two-way fixed-effects model.

\section*{Declarations}

\textbf{Availability of data and materials:}
The patent data used in this study were obtained from the Japan Patent Office via the J-platPat portal (\url{https://www.j-platpat.inpit.go.jp/}).
Gross regional product and population data were obtained from the Statistics Bureau of Japan via the e-Stat portal (\url{https://www.e-stat.go.jp}).
Patent classification concordance tables mapping International Patent Classification codes to Schmoch technology fields were acquired from the World Intellectual Property Organization IP Statistics portal (\url{https://www.wipo.int/en/web/ip-statistics}).

\textbf{Code availability:}
All scripts and code used to compute the Fitness index and reproduce the results of this paper are publicly available at \url{https://github.com/RintaroKARASHIMA/KCIinJapaneseFirms}.

\textbf{Competing interests:}
The authors declare no competing interests.

\textbf{Funding:}
This work was supported by Japan Society for the Promotion of Science KAKENHI Grant-in-Aid for Scientific Research (B) Grant Numbers JP25K01454 and JP24K00247, and Watanabe Memorial Foundation for the Advancement of New Technology FY2024 Science and Technology Research Grant Number R6-590.
The funders had no role in study design, data collection and analysis, decision to publish, or preparation of the manuscript.

\textbf{Authors' contributions:}
H.I.\ conceived the main research idea, designed the study, collected the data, and secured the funding.
R.K.\ performed the analysis and drafted the manuscript.
Both H.I.\ and R.K.\ critically reviewed the results, discussed the findings, contributed to the revisions of the manuscript, and approved the final version of the manuscript.

\textbf{Acknowledgements:}
This work was supported by Japan Society for the Promotion of Science KAKENHI Grant-in-Aid for Scientific Research (B) Grant Numbers JP25K01454 and JP24K00247, and Watanabe Memorial Foundation for the Advancement of New Technology FY2024 Science and Technology Research Grant Number R6-590.
The funders had no role in study design, data collection and analysis, decision to publish, or preparation of the manuscript.

\begin{appendices}

\section{Supplementary Information}\label{secA1}

\renewcommand{\thefigure}{S\arabic{figure}}
\renewcommand{\thetable}{S\arabic{table}}
\setcounter{figure}{0}
\setcounter{table}{0}

\begin{figure}[ht]
  \includegraphics[scale=0.8]{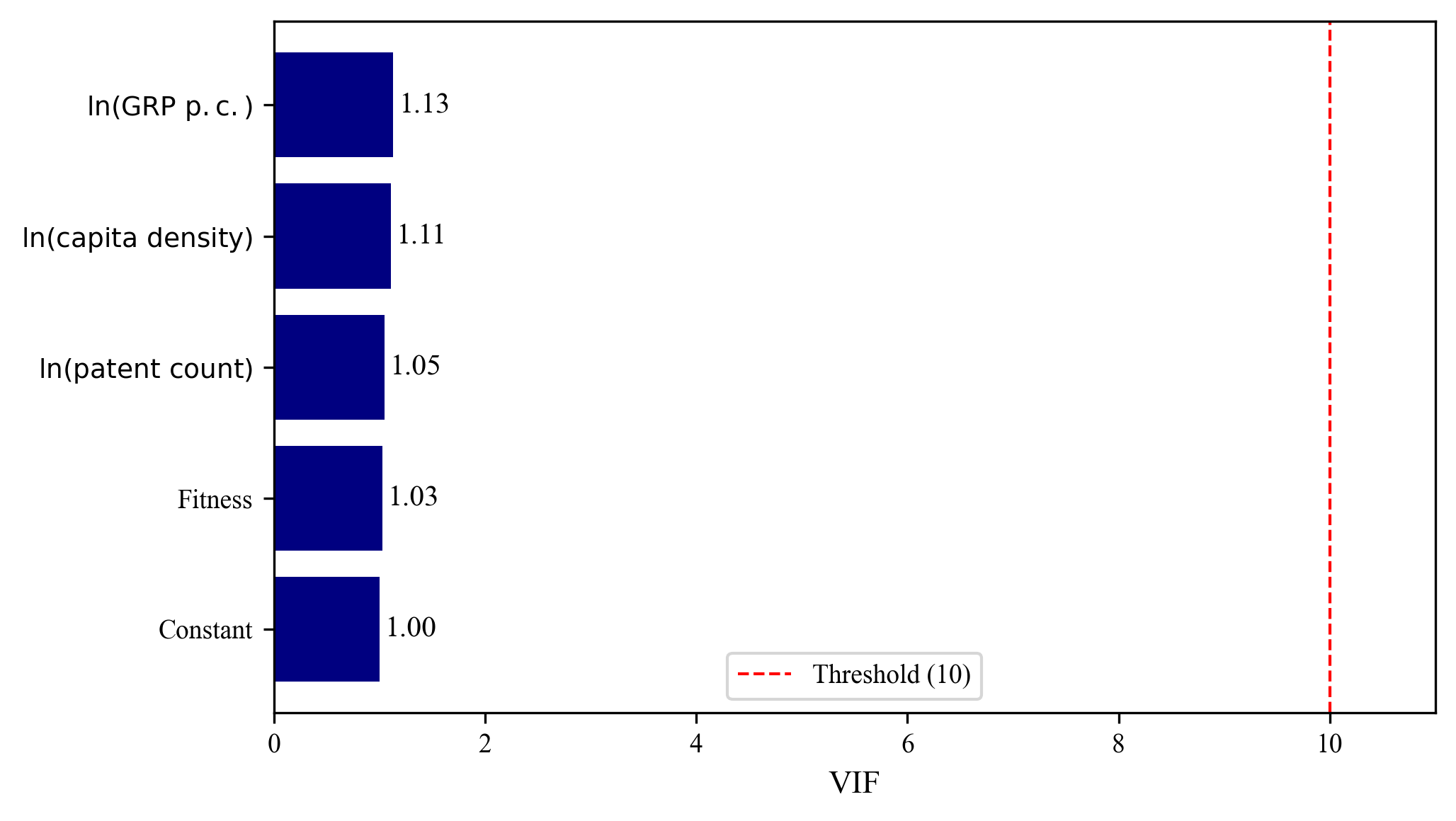}
  \caption{%
    Variance inflation factors (VIF) for all regressors after applying the within-group transformation of the two-way fixed-effects model.
    All values fall below 1.15, well below the conventional threshold of 10, confirming that multicollinearity is not a concern.%
  }
  \label{fig:vif}
\end{figure}

\begin{figure}[ht]
  \includegraphics[scale=0.6]{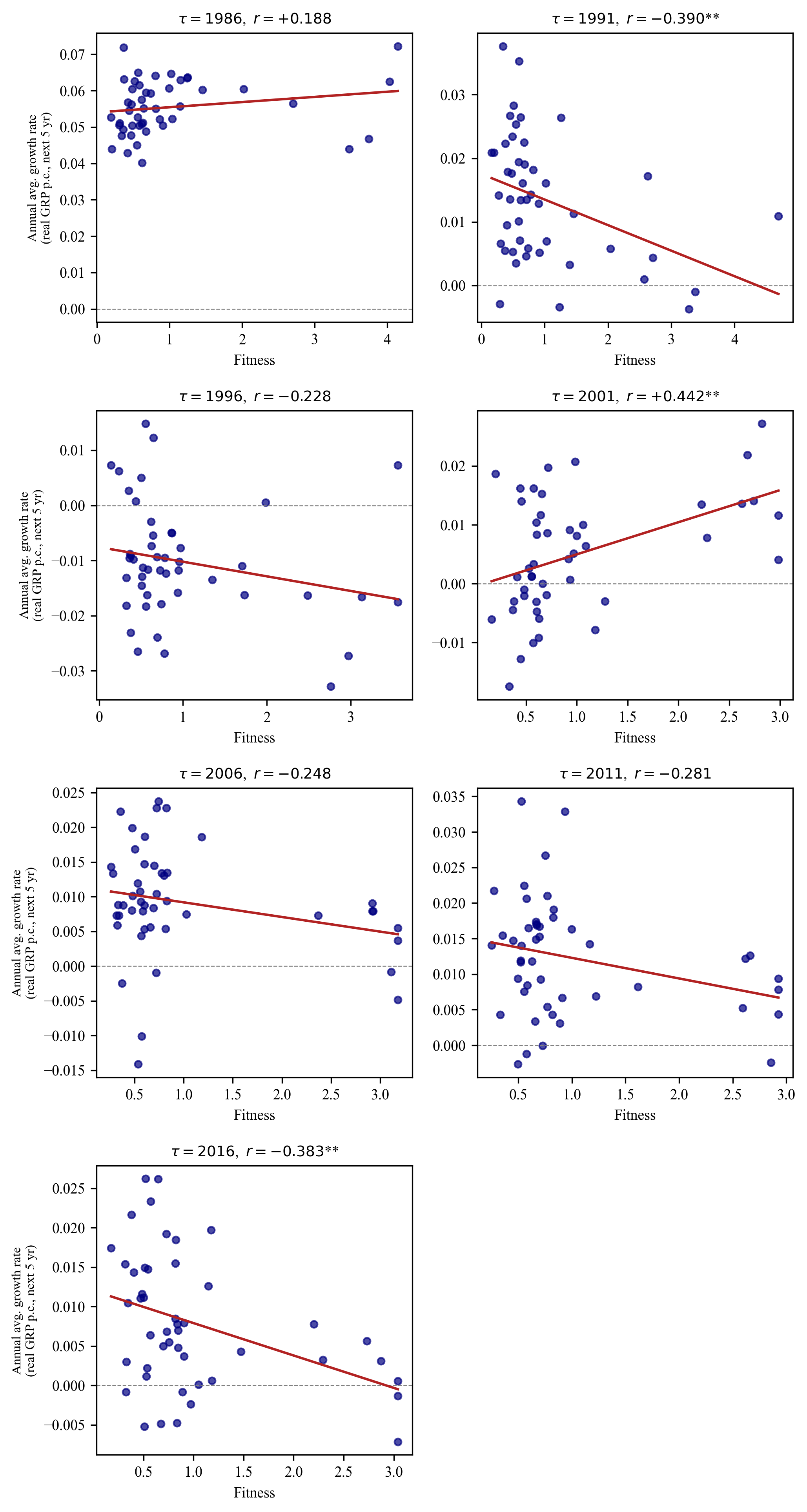}
  \caption{%
    Scatter plots of prefectural Fitness (horizontal axis) against the subsequent five-year annual average growth rate of real GRP per capita (vertical axis) for each of the seven periods.
    The sign and significance of the cross-sectional correlation varies across periods, ranging from positive and insignificant in 1986 to negative and significant in 1991 and 2016.%
  }
  \label{fig:crosssectional}
\end{figure}

\begin{table}[ht]
\caption{Full coefficient estimates for the two-way fixed-effects model (Model 4)}\label{tab:fullmodel}
\begin{tabular}{lcccc}
\toprule
Variable & Coefficient & Std.\ Error & $t$-statistic & $p$-value \\
\midrule
Intercept                  &  2.034 & 0.397 &  5.12 & $<$0.001$^{***}$ \\
$\ln(\text{GRP pc})$       & $-0.113$ & 0.024 & $-4.69$ & $<$0.001$^{***}$ \\
Fitness                    &  0.003 & 0.001 &  2.72 & $0.007^{**}$     \\
$\ln(\text{capita density})$ & $-0.059$ & 0.018 & $-3.31$ & $0.001^{***}$  \\
$\ln(\text{patent count})$ &  0.003 & 0.001 &  3.12 & $0.002^{**}$     \\
\midrule
Within $R^2$ & \multicolumn{4}{c}{0.694} \\
$N$          & \multicolumn{4}{c}{329}   \\
\bottomrule
\multicolumn{5}{l}{Driscoll-Kraay standard errors with bandwidth = 3, with entity and time fixed effects.}\\
\multicolumn{5}{l}{$^{**}p < 0.01$; $^{***}p < 0.001$.}\\
\end{tabular}
\end{table}

\end{appendices}

\bibliography{sn-bibliography}

@article{hidalgo2009eci,
    author = {C{\'e}sar A. Hidalgo and Ricardo Hausmann },
    title = {The building blocks of economic complexity},
    journal = {Proceedings of the National Academy of Sciences},
    volume = {106},
    number = {26},
    pages = {10570-10575},
    year = {2009},
    doi = {10.1073/pnas.0900943106},
    memo = {HH algorithmを提案}
}

@article{Hidalgo2021,
  title={Economic complexity theory and applications},
  author={Hidalgo, C{\'e}sar A},
  journal={Nature Reviews Physics},
  volume={3},
  number={2},
  pages={92--113},
  year={2021},
  publisher={Nature Publishing Group UK London},
  memo = {HH algorithmの応用をまとめたsurvey}
}

@article{balland2017tci,
    author = {Pierre-Alexandre Balland and David Rigby},
    title = {The Geography of Complex Knowledge},
    journal = {Economic Geography},
    volume = {93},
    number = {1},
    pages = {1--23},
    year = {2017},
    publisher = {Routledge},
    doi = {10.1080/00130095.2016.1205947},
    memo = {TCIを提案}
}

@ARTICLE{Jun2023,
  author={Jun, Bogang and Kim, Seung Hwan and Choi, Hyoji and Jeon, Jeong Hwan and Yu, Donghyeon},
  journal={IEEE Access},
  title={Technological Leadership in Industry 4.0: A Comparison Between Manufacturing and ICT Sectors Among Korean Firms},
  year={2023},
  volume={11},
  number={},
  pages={28490-28505},
  doi={10.1109/ACCESS.2023.3259065},
  ISSN={2169-3536},
  month={},
  memo = {TCIを企業レベルに応用}
}

@article{ballandandboschma2021,
  author = {Pierre-Alexandre Balland and Ron Boschma},
  title = {Mapping the potentials of regions in Europe to contribute to new knowledge production in Industry 4.0 technologies},
  journal = {Regional Studies},
  volume = {55},
  number = {10-11},
  pages = {1652--1666},
  year = {2021},
  publisher = {Routledge},
  doi = {10.1080/00343404.2021.1900557},
  URL = {https://doi.org/10.1080/00343404.2021.1900557},
  eprint = {https://doi.org/10.1080/00343404.2021.1900557}
}

@article{Hidalgo2023,
  author  = {Hidalgo, C. A.},
  title   = {The policy implications of economic complexity},
  journal = {Research Policy},
  volume  = {52},
  number  = {9},
  pages   = {104863},
  year    = {2023}
}

@article{Balland2022,
  author  = {Balland, P. A. and Broekel, T. and Diodato, D. and Giuliani, E. and Hausmann, R. and O'Clery, N. and Rigby, D.},
  title   = {Reprint of the new paradigm of economic complexity},
  journal = {Research Policy},
  volume  = {51},
  number  = {8},
  pages   = {104568},
  year    = {2022}
}

@article{Balland2018,
  author  = {Balland, P. A. and Boschma, R. and Crespo, J. and Rigby, D. L.},
  title   = {Smart specialization policy in the European Union: relatedness, knowledge complexity and regional diversification},
  journal = {Regional Studies},
  volume  = {53},
  number  = {9},
  pages   = {1252--1268},
  year    = {2018}
}

@article{PintarScherngell2022,
  author  = {Pintar, N. and Scherngell, T.},
  title   = {The complex nature of regional knowledge production: Evidence on European regions},
  journal = {Research Policy},
  volume  = {51},
  number  = {8},
  pages   = {104170},
  year    = {2022}
}

@article{Pinheiro2022,
  author  = {Pinheiro, F. L. and Balland, P. A. and Boschma, R. and Hartmann, D.},
  title   = {The dark side of the geography of innovation: relatedness, complexity and regional inequality in Europe},
  journal = {Regional Studies},
  pages   = {1--16},
  year    = {2022},
  number = {0},
  publisher = {Routledge},
  doi = {10.1080/00343404.2022.2106362},
  URL = {https://doi.org/10.1080/00343404.2022.2106362},
  eprint = {https://doi.org/10.1080/00343404.2022.2106362}
}

@article{Whittle2019,
  author  = {Whittle, A.},
  title   = {Local and nonlocal knowledge typologies: technological complexity in the Irish knowledge space},
  journal = {European Planning Studies},
  volume  = {27},
  number  = {4},
  pages   = {661--677},
  year    = {2019}
}

@article{Abay2024,
  author  = {Abay, M. and Akg{\"u}ng{\"o}r, S.},
  title   = {Technological paths and smart specialization: analysis of regional entry and exit in Turkey},
  journal = {Asia-Pacific Journal of Regional Science},
  volume  = {8},
  number  = {1},
  pages   = {45--84},
  year    = {2024}
}

@article{Chakraborty2020,
  author  = {Chakraborty, A. and Inoue, H. and Fujiwara, Y.},
  title   = {Economic complexity of prefectures in Japan},
  journal = {PloS One},
  volume  = {15},
  number  = {8},
  pages   = {e0238017},
  year    = {2020}
}

@techreport{Schmoch2008,
  author      = {Schmoch, U.},
  title       = {Concept of a technology classification for country comparisons},
  institution = {World Intellectual Property Organisation (WIPO)},
  year        = {2008},
  note        = {Final report}
}

@article{Soete1987,
  author  = {Soete, L.},
  title   = {The impact of technological innovation on international trade patterns: the evidence reconsidered},
  journal = {Research Policy},
  volume  = {16},
  number  = {2--4},
  pages   = {101--130},
  year    = {1987}
}

@article{Balassa1965,
  author  = {Balassa, B.},
  title   = {Trade Liberalisation and ``Revealed'' Comparative Advantage},
  journal = {The Manchester School},
  volume  = {33},
  number  = {2},
  pages   = {99--123},
  year    = {1965}
}

@misc{PintarEssletzbichler2022,
  author       = {Pintar, N. and Essletzbichler, J.},
  title        = {Complexity and smart specialization: Comparing and evaluating knowledge complexity measures for European city-regions},
  howpublished = {Preprint of Papers in Economic Geography and Innovation Studies},
  year         = {2022}
}

@article{Tacchella2012,
  author  = {Tacchella, A. and Cristelli, M. and Caldarelli, G. and Gabrielli, A. and Pietronero, L.},
  title   = {A new metrics for countries' fitness and products' complexity},
  journal = {Scientific Reports},
  volume  = {2},
  number  = {1},
  pages   = {723},
  year    = {2012}
}

@article{Albeaik2017,
  author    = {Albeaik, S. and Kaltenberg, M. and Alsaleh, M. and Hidalgo, C. A.},
  title     = {Improving the economic complexity index},
  journal   = {arXiv preprint},
  eprint    = {arXiv:1707.05826},
  year      = {2017}
}

@article{morrison2017economic,
  title={On economic complexity and the fitness of nations},
  author={Morrison, Greg and Buldyrev, Sergey V and Imbruno, Michele and Doria Arrieta, Omar Alonso and Rungi, Armando and Riccaboni, Massimo and Pammolli, Fabio},
  journal={Scientific reports},
  volume={7},
  number={1},
  pages={15332},
  year={2017},
  publisher={Nature Publishing Group UK London}
}

@article{straccamore2023urban,
  title={Urban economic fitness and complexity from patent data},
  author={Straccamore, Matteo and Bruno, Matteo and Monechi, Bernardo and Loreto, Vittorio},
  journal={Scientific Reports},
  volume={13},
  number={1},
  pages={3655},
  year={2023},
  publisher={Nature Publishing Group UK London}
}

@article{straccamore2025comparative,
  title={Comparative analysis of technological fitness and coherence at different geographical scales},
  author={Straccamore, Matteo and Bruno, Matteo and Tacchella, Andrea},
  journal={PLoS One},
  volume={20},
  number={8},
  pages={e0329746},
  year={2025},
  publisher={Public Library of Science San Francisco, CA USA}
}

@article{mewes2022technological,
  title={Technological complexity and economic growth of regions},
  author={Mewes, Lars and Broekel, Tom},
  journal={Research Policy},
  volume={51},
  number={8},
  pages={104156},
  year={2022},
  publisher={Elsevier}
}

@article{FritzManduca2021,
  author  = {Fritz, B. S. and Manduca, R. A.},
  title   = {The economic complexity of {US} metropolitan areas},
  journal = {Regional Studies},
  volume  = {55},
  number  = {7},
  pages   = {1299--1310},
  year    = {2021}
}

@article{DriscollandKraay1998,
  author  = {Driscoll, J. C. and Kraay, A. C.},
  title   = {Consistent covariance matrix estimation with spatially dependent panel data},
  journal = {Review of Economics and Statistics},
  volume  = {80},
  number  = {4},
  pages   = {549--560},
  year    = {1998}
}

@article{Pugliese2016,
  author  = {Pugliese, E. and Zaccaria, A. and Pietronero, L.},
  title   = {On the convergence of the fitness-complexity algorithm},
  journal = {The European Physical Journal Special Topics},
  volume  = {225},
  number  = {10},
  pages   = {1893--1911},
  year    = {2016}
}

@article{LiRigby2023,
  author  = {Li, Y. and Rigby, D.},
  title   = {Relatedness, complexity, and economic growth in {C}hinese cities},
  journal = {International Regional Science Review},
  volume  = {46},
  number  = {1},
  pages   = {3--37},
  year    = {2023}
}

@article{Bahrami2023,
  author  = {Bahrami, F. and Shahmoradi, B. and Noori, J. and Turkina, E. and Bahrami, H.},
  title   = {Economic complexity and the dynamics of regional competitiveness: a systematic review},
  journal = {Competitiveness Review: An International Business Journal},
  volume  = {33},
  number  = {4},
  pages   = {711--744},
  year    = {2023}
}

\end{document}